\documentclass{elsart5p}
\usepackage[numbers,sort&compress]{natbib}
\usepackage{amsmath}
\usepackage{amssymb}
\usepackage{latexsym}
\usepackage[dvips]{graphicx}
\usepackage[dvips]{epsfig}
\newcommand{\beq}{\begin{equation}}
\newcommand{\eeq}{\end{equation}}

\newcommand{\dgr}{ {\,}^{\circ} \mbox{C}}
\newcommand{\un}[1]{\ensuremath{\unskip\,\mathrm{#1}}}

\begin{document}

\begin{frontmatter}
\title{Membrane-Mediated Repulsion between Gramicidin Pores}
\author{Doru Constantin\ead{constantin@lps.u-psud.fr}}
\address{Laboratoire de Physique des Solides, Universit\'e Paris-Sud,
CNRS, UMR 8502, F-91405 Orsay Cedex, France.}

\begin{abstract}
We investigated the X-ray scattering signal of highly aligned
multilayers of the zwitterionic lipid
1,2-dilauroyl-sn-glycero-3-phosphatidylcholine containing pores
formed by the antimicrobial peptide gramicidin as a function of the
peptide/lipid ratio. We are able to obtain information on the
structure factor of the pore fluid, which then yields the
interaction potential between pores in the plane of the bilayers.
Aside from a hard core with a radius close to the geometric radius
of the pore, we find a repulsive exponential lipid-mediated
interaction with a decay length of {2.5~\AA} and an amplitude that
decreases with the pore concentration, in agreement with the
hydrophobic matching hypothesis. In dilute systems, the contact
value of this interaction is about $30 \, k_B T$. Similar results
are obtained for gramicidin pores inserted within bilayers formed by
the nonionic surfactant pentaethylene glycol monododecyl ether.
\end{abstract}

\begin{keyword}
Gramicidin \sep Small-angle x-ray scattering \sep In-plane
interaction \sep Hydrophobic mismatch

\PACS 61.05.cf \sep 87.16.dt \sep 87.15.kt

\end{keyword}
\end{frontmatter}

\section{Introduction}\label{sec:intro}

In the last decades, much effort has been dedicated to the
understanding of biological membranes, in particular to the
interaction between membrane proteins and the host lipid bilayer.
While the fluid mosaic model \cite{Singer:1972} described the
proteins as free to diffuse in this environment, it was perceived
very early on \cite{Sackmann:1984} that, far from being a neutral
background, the lipid bilayer can influence protein organization in
the plane of the membrane and hence many aspects of their activity
(such as cell signaling and membrane trafficking). In particular,
cholesterol content was recognized as an important variable. For
instance, it was put forward as a regulating parameter for protein
partitioning between the plasma membrane and the Golgi complex
\cite{Bretscher:1993}. More generally, the homogeneity of
cholesterol-containing membranes and the biological relevance of the
so-called lipid rafts \cite{Simons:1997} has been the subject of
extensive research \cite{Simons:2000}.

Interestingly enough, the mechanisms involved in protein-lipid
interaction and in the resulting protein-protein interaction
mediated by the lipid membrane are non-specific, in that no chemical
bond is formed between a protein and a small number of definite
lipid molecules. Thus, a detailed understanding and a quantitative
characterization of the interaction between lipids and proteins in
membranes require that one considers the membrane as a many-particle
system whose properties are collectively determined by the assembly
and not only by the chemical properties of the individual lipids and
proteins \cite{Jensen:2004}. This justifies the hope that
--notwithstanding the complexity of the system-- the concepts
developed in soft matter physics for the understanding of
self-assembled systems are operative in this context and that
'simplified' models can yield valuable information. For this reason,
a considerable body of work dealing with the theoretical modelling
and numerical simulation of protein-lipid systems appeared in the
last decades, see \cite{Sperotto:2006} for a recent review. These
efforts are either continuum-elasticity theories or more detailed
models taking into account the molecular structure of the lipid
bilayer.

However, very few experiments attempted to determine directly the
interaction forces between membrane inclusions. First among them,
freeze-fracture electron microscopy (FFEM) studies
\citep{Lewis:1983,Chen:1973,James:1973,Abney:1987} yielded the
radial distribution function of the inclusions. Comparing the data
to liquid state theories
\citep{Pearson:1983,Pearson:1984,Braun:1987} resulted in a hard-core
model with, in some cases, an additional repulsive or attractive
interaction. FFEM was not extensively used, undoubtedly due to the
inherent experimental difficulties; moreover, it is not obvious that
the distribution measured in the frozen sample is identical to that
at thermal equilibrium.

The interaction of membrane inclusions can also be studied using
small-angle neutron or X-ray scattering from oriented samples, as
demonstrated by Huang and collaborators
\citep{He:1995,He:1996,Yang:1999}. One can thus measure the
structure factor of the two-dimensional system formed by the
inclusions in the membrane. Further analysis gives access to the
interaction potential of the inclusions. In our experience, this
method is more reliable when several measurements are performed
along a dilution line (varying density of inclusions) and the
results are treated simultaneously. As an application, we could
measure a repulsive interaction both for alamethicin pores in DMPC
membranes \citep{Constantin:2007} and for inorganic particles in
synthetic membranes \citep{Constantin:2008}.

In this paper, we study pores formed by the antimicrobial peptide
gramicidin D in bilayers with different compositions. Simultaneously
fitting the two-dimensional structure factor of the pores in the
plane of the membrane yields the interaction potential between the
pores.

\section{Materials and Methods}\label{sec:matmeth}

\subsection{Sample preparation and environment}
\label{subsec:sample_prep}

The lipid 1,2-dilauroyl-sn-glycero-3-phos\-pha\-ti\-dyl\-cho\-line
(DLPC) was purchased from Avanti Polar Lipids Inc. (Birmingham, AL,
USA). The antimicrobial peptide gramicidin D and the zwitterionic
surfactant N,N-dimethyldodecylamine-N-oxide (DDAO) were bought from
Sigma Aldrich. The nonionic surfactant pentaethylene glycol
monododecyl ether (C$_{12}$EO$_{5}$) was bought from Nikko Chemical
Ltd. (Japan). Without further purification, the products were
dissolved in isopropanol. The stock solutions were then mixed to
give the desired molar peptide/lipid ratio $P/L$. The resulting
solutions were then dried in vacuum and hydrated in excess water
(DLPC phases) or up to a water content of 20~wt\% (DDAO and
C$_{12}$EO$_{5}$).

The samples were prepared in flat glass capillaries (VitroCom Inc.,
Mt. Lks, NJ, USA), 100~$\mu$m thick and 2~mm wide by gently sucking
in the lamellar phase using a syringe. The capillaries were
flame-sealed. Areas aligned in homeotropic alignment (lamellae
parallel to the flat faces of the capillary) formed slowly (over a
few months, at room temperature) in DLPC phases. Samples of DDAO and
C$_{12}$EO$_{5}$ were aligned by thermal cycling between the
lamellar and isotropic phases, at cooling rates of about $1\dgr /
\un{min}$.

\subsection{Measurement}

The SAXS measurements were performed at the bending magnet beamline
BM02 (D2AM) of the European Synchrotron Radiation Facility (ESRF,
Grenoble, France). The photon energy was set at 11 keV. See
reference \cite{Simon:1997} for more details.

The data was acquired using a Peltier-cooled CCD camera (SCX90-1300,
from Princeton Instruments Inc., NJ, USA) with a resolution of $1340
\times 1300$ pixels. Data preprocessing (dark current subtraction,
flat field correction, radial regrouping and normalization) was
performed using the \texttt{bm2img} software developed at the
beamline.

The incident beam was perpendicular to the flat face of the
capillary (parallel to the smectic director, which we take along the
$z$ axis.) Thus, the scattering vector $\mathbf{q}$ is mostly
contained in the $(x,y)$ plane of the layers, and the measured
scattered signal $I(\mathbf{q})$ probes inhomogeneities of the
electron density in this plane. Since the bilayers form a
two-dimensional liquid, the scattering pattern exhibits azimuthal
symmetry: $I=I(q=|\mathbf{q}|)$. The capillaries were scanned in the
beam to find well-aligned domains (where the intensity of the
residual Bragg reflections was as low as possible.)

\subsection{Analysis}

The gramicidin pores are dispersed in the lamellar phase matrix.
Since the ``pure'' lamellar phase gives a signal confined to the
vicinity of the Bragg peaks, from the Babinet principle it ensues
that the off-axis scattering is the same as for a system where the
density profile of the lamellar phase is subtracted, and one is left
with fictitious ``pore minus bilayer'' objects in a completely
transparent medium. Furthermore, as the pores represent a collection
of identical and similarly oriented objects (up to an azimuthal
averaging), the classical separation of the scattering intensity in
a structure factor multiplied by a form factor can be applied
\citep{Chaikin:1995}, yielding: $I(\mathbf{q})=S(\mathbf{q})\cdot
Ff(\mathbf{q})$, with:

\beq \label{eq:struct} S(q_z,q_r)= \frac{1}{N} \left \langle \left |
\sum_{k=1}^{N-1} \exp \left ( - i \mathbf{q} \mathbf{r}_k \right )
\right | ^2 \right \rangle \eeq

\noindent where $N$ is the number of objects and object ``0'' is
taken as the origin of the coordinates. If there is no in-plane
ordering, $S$ only depends on the absolute value of the in-plane
scattering vector $q_r = \sqrt{q_x^2 + q_y^2}$.

\begin{figure}
\includegraphics[width=0.5\textwidth]{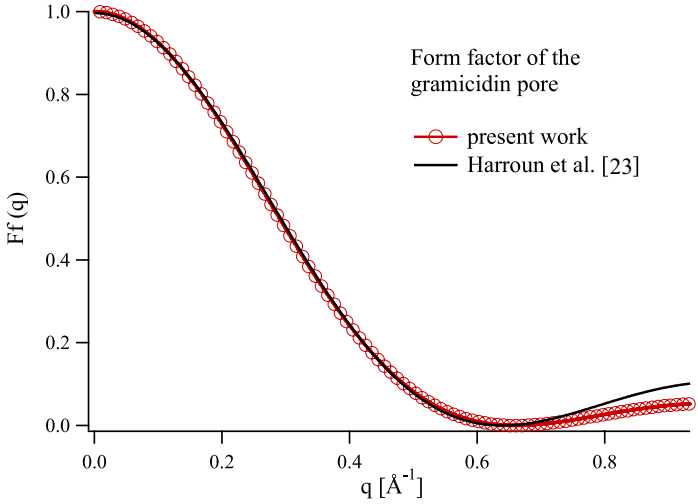}
\caption{Normalized in-plane form factor of the gramicidin pore
(helical dimer configuration) obtained from the atomic configuration
of deGroot et al. \cite{deGroot:2002} (line and symbols) compared to
the form factor used by Harroun et al. \cite{Harroun:1999} (solid
line).} \label{fig:form_factor}
\end{figure}

The form factor $Ff(q_r)$ is the squared modulus of the Fourier
transform of the electron density $\rho(r)$ of the scattering
object. We computed it from the atomic coordinates of the starting
structure used in the molecular dynamics (MD) simulation of
\cite{deGroot:2002}, projected onto the $z$ plane. The main feature
of the density profile is the increased electron density at the
position of the backbone, which we fitted by a radial Gaussian
profile, with a peak radius $r_0 = 3.47 \un{\AA}$ and a width $w=1.5
\un{\AA}$. The resulting from factor (Figure~\ref{fig:form_factor})
is very close to that obtained by \cite{Harroun:1999} from the MD
simulations of \cite{Woolf:1996}.

\begin{figure}
\includegraphics[width=0.5\textwidth]{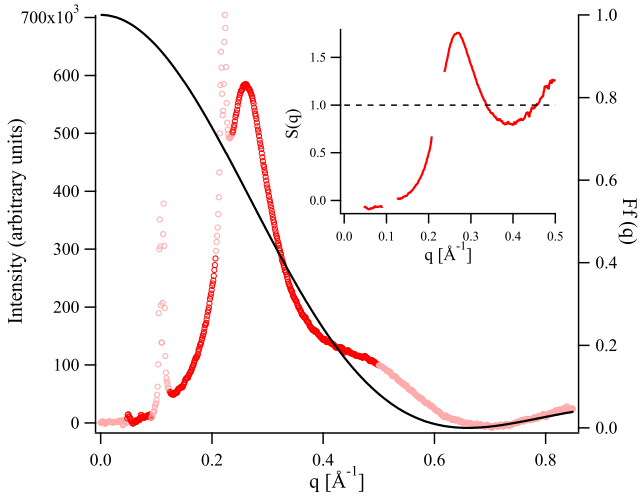}
\caption{Scattered intensity $I(q)$ (symbols) for a sample with
$P/L=1/7.5$ and normalized form factor (solid line); see
Figure~\ref{fig:form_factor}. Inset: the structure factor obtained
by dividing the measured intensity through the form factor. The
light symbols in the intensity curve (and gaps in the structure
factor) correspond to unusable data: at small angles (in the shade
of the beamstop), at the position of the lamellar peaks and at high
angles, where a systematic discrepancy is observed between the
intensity and the form factor.} \label{fig:Iq}
\end{figure}

The intensity is divided by the form factor to yield the
two-dimensional structure factor $S(q)$ of the fluid formed by the
pores in the plane of the membrane (Figure~\ref{fig:Iq}.) The
uncertainty in the intensity is estimated from the spread of the
values recorded by the different pixels with the same $q$ value. We
use standard error propagation through the background subtraction
and form factor division steps to estimate the uncertainty in the
structure factor values.

The sample alignment is not perfect and sometimes residual lamellar
peaks persist, as seen in Figure~\ref{fig:Iq}. The points around
these positions are discarded (light symbols in the $I(q)$ curve and
gaps in the $S(q)$ curve.) We also discard the points at low
$q$-values (in the shade of the beamstop) and those above
$0.5~\un{\AA}^{-1}$, where a systematic discrepancy between the
experimental data and the model form factor leads to an artificial
oscillation of the structure factor. In spite of these limitations,
the first peak of the structure factor is properly measured for all
samples.

\section{Results and Discussion}

The structure factor curves are plotted as open dots in
Figure~\ref{fig:HD_fit}~a) for all $P/L$ values, indicated alongside
the curves. The uncertainty bars are shown for all data points (they
are generally smaller than the symbol size.) In the following, we
will make the simplifying assumption that there is no interaction
between the pores along $z$ (from one bilayer to the next), a result
obtained by \cite{Yang:1999} for gramicidin pores in fully hydrated
bilayers. This feature is extremely important for two reasons: first
of all, it allows us to use a two-dimensional model, only dealing
with the interaction within a bilayer; second, it ensures the
biological relevance of this study.

\subsection{Hard disk model}

\begin{figure*}
\includegraphics[width=\textwidth]{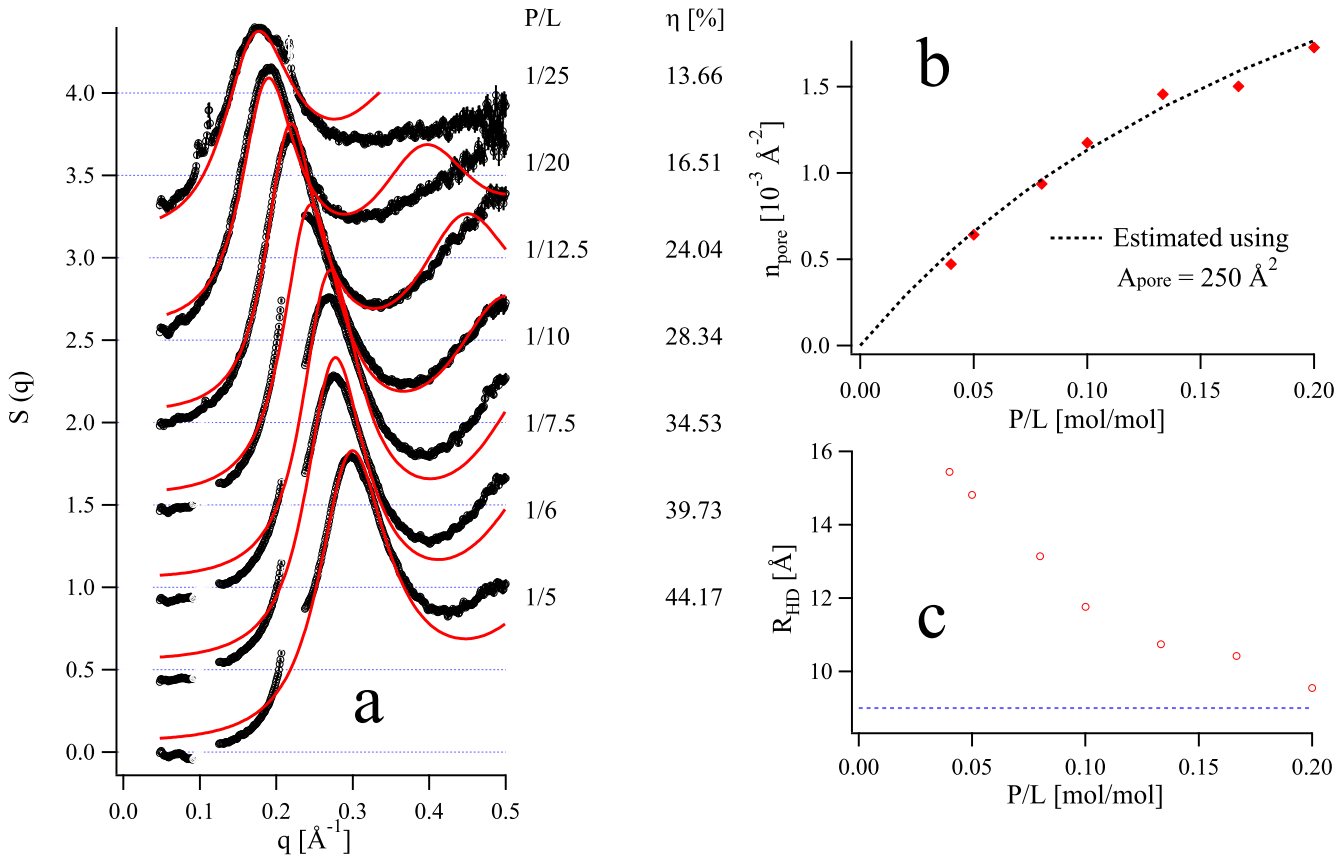}
\caption{(a) Experimental structure factors (symbols) and hard-disk
fits (solid lines) for the gramicidin/DLPC system, at different
peptide/lipid (P/L) molar concentrations indicated alongside the
curves. The theoretical surface fraction of gramicidin pores,
$\eta$, is also given. (b) Pore density obtained from the fits in
(a) (symbols) and theoretically expected value (dotted line). (c)
Effective values for the hard disk radius obtained from the fits in
(a) (symbols) and geometrical pore radius (dotted line). The
variation in $R_{HD}$ with the pore concentration is a sign of
repulsive interaction, as discussed below.} \label{fig:HD_fit}
\end{figure*}

The simplest model for the interaction of gramicidin pores in
membranes is that of hard disks confined in the plane. Such an
analysis was already performed by \cite{Harroun:1999} for gramicidin
in DLPC bilayers at $P/L=1/10$. As a first step, we analysed all the
curves using the two-dimensional structure factor
$S_{\mathrm{hd}}(q_r)$, given by the simple analytical expression
obtained by \cite{Rosenfeld:1990} (see Eq. 6.8) using the
``fundamental measure'' approach. The details are shown in
Figure~\ref{fig:HD_fit} for the gramicidin/DLPC system: panel a)
displays the data and fits, while panels b) and c) show the
evolution with $P/L$ of the fit parameters, namely the number
density of the pores $n_{\rm pore}$ and the effective hard-disk
radius $R_{HD}$. Both parameters vary freely during the fit.

A first observation is that the pore density (symbols in
Figure~\ref{fig:HD_fit}~b) is in very good agreement with the
theoretical value calculated using the molar ratio $P/L$ and
published data for the area per lipid/surfactant molecule and
gramicidin pore. A similar agreement is obtained for the two other
systems: gramicidin/C$_{12}$EO$_{5}$ and gramicidin/DDAO (data not
shown).

\begin{figure}
\includegraphics[width=0.5\textwidth]{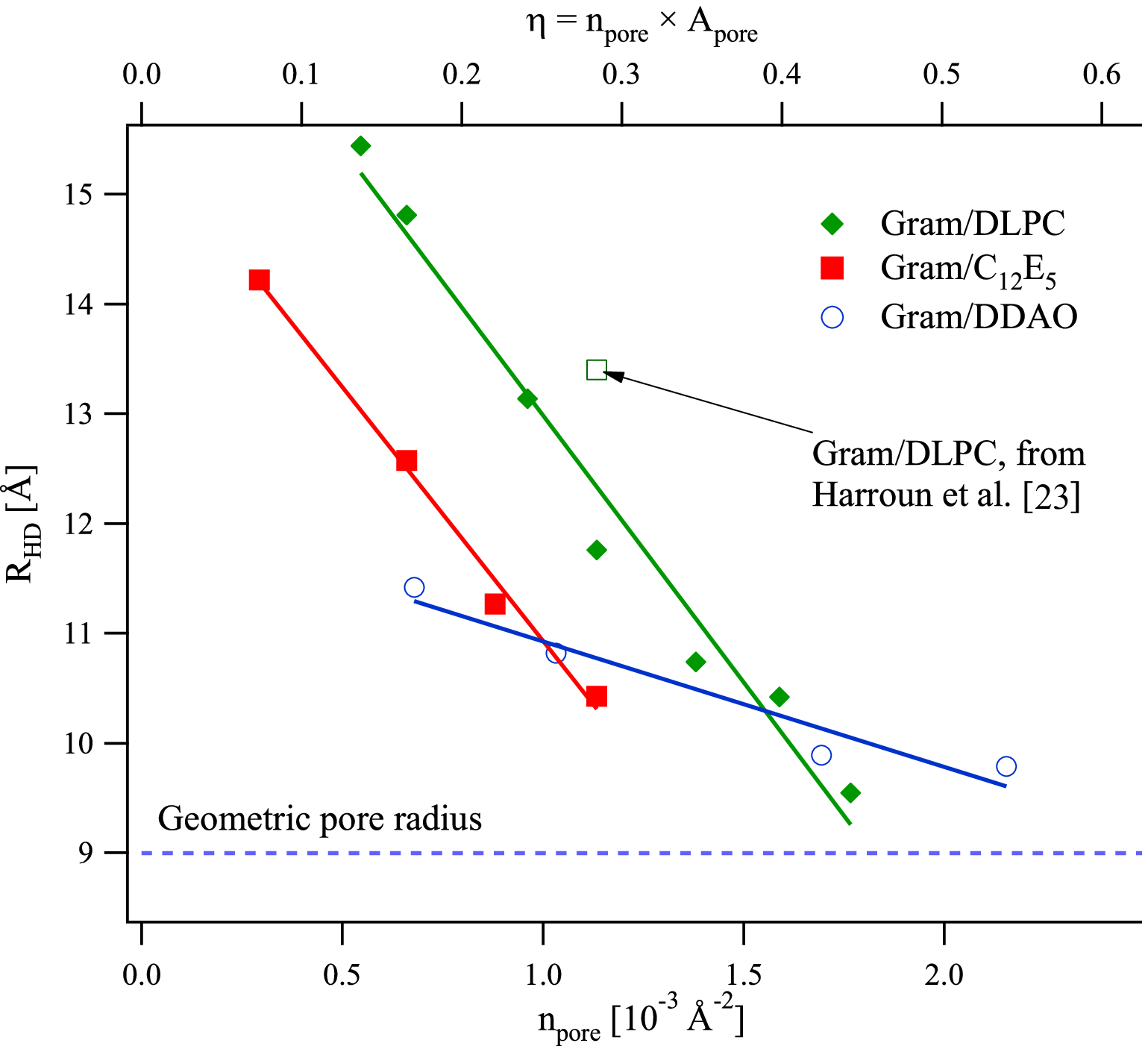}
\caption{Effective hard disk radius of the gramicidin pore as a
function of the density in bilayers with three different
compositions. The lines are just guides for the eye.}
\label{fig:HDradius}
\end{figure}

The second fit parameter, the effective hard disk radius $R_{HD}$ is
shown as a function of $P/L$ in Figure~\ref{fig:HD_fit}~c) and as a
function of the pore density $n_{\rm pore}$ and the area fraction
$\eta = n_{\rm pore} \times A_{\rm pore}$ (with $A_{\rm pore}=250 \,
\un{\AA}^2$ the area of a gramicidin pore) in
Figure~\ref{fig:HDradius}. For comparison, we also plot the data
point of \cite{Harroun:1999} (open square). From the analysis of the
structure factor for Gram/DLPC at $P/L=1/10$, these authors obtain a
hard-disk radius of 13.4~{\AA}, which they interpret as the
geometric radius of the pore plus one lipid layer. This value is
somewhat higher, but still compatible with our data\footnote{This
discrepancy might stem from the slightly different way of obtaining
the interaction radius. Harroun et al. \cite{Harroun:1999} determine
the most probable nearest-neighbor separation between two pores,
always larger than twice the hard-disk radius, which is the
\textit{minimum} separation.}.

In Figure~\ref{fig:HDradius}, we also show the fit results for the
two other systems. For all three membrane compositions, the
effective radius decreases as a function of $n_{\rm pore}$ (although
this decrease is less marked for gramicidin/DDAO.) Clearly, the hard
disk model is not satisfactory: although it fits very well the
individual curves, the interaction radius decreases as $P/L$ (and
hence $n_{\rm pore}$) increase; as already discussed for the case of
alamethicin pores \citep{Constantin:2007}, this is a sign of an
additional ``soft'' repulsive interaction. Briefly, this effect can
be understood as follows: at low density, this interaction is enough
to keep the particles far from each other. As the density increases
so does the pressure, and the particles are eventually pushed closer
and closer to each other. At high concentration, the effective
radius should saturate at the (impenetrable) ``true'' core value.
This seems indeed to be the case for the data in
Figure~\ref{fig:HDradius}, where at increasing density the curves
approach the geometric core radius of 9~{\AA}.

\subsection{Additional interaction}

To quantify this tendency, we calculated the structure factors for a
hard core with radius $R_{HD} = 9 \un{\AA}$ and an additional
``soft'' potential:

\begin{equation}
V(r) = u \exp \left [ -\frac{1}{2} \left ( \frac{r}{\xi} \right ) ^2
\right ] \quad r
> 2 R_{HD}
\label{eq:vr}
\end{equation}
where $r$ denotes the distance between the pore centres.

The structure factor $S(q)$ is now a function of four parameters:
the hard core radius $R_{HD}$, the number density $n_{\rm pore}$, as
well as the amplitude $u$ and the decay length $\xi$ of the
additional component. We calculate $S(q)$ using the method of Lado
\cite{Lado:1967,Lado:1968}, implemented as an \textsc{Igor Pro}
function\footnote{The source code is available from the author upon
request.}. Briefly, the method provides an iterative solution to the
Ornstein-Zernicke equation with the Percus-Yevick closure. The
accuracy of the procedure was tested by comparing the results for
purely hard-core systems to the analytical formula of Rosenfeld
\citep{Rosenfeld:1990}.

\begin{figure}
\includegraphics[width=0.5\textwidth]{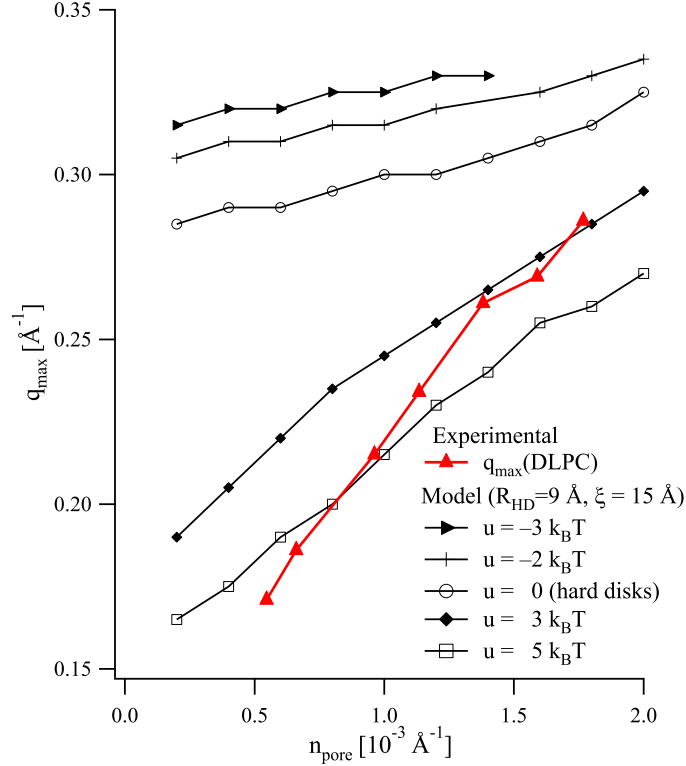}
\caption{Position of the first maximum $q_{\rm max}$ in the
structure factor $S(q)$. Experimental data ($\blacktriangle$ and
thick line) for the Gram/DLPC system and simulated data (various
symbols and lines) for different amplitudes of the additional
interaction $u$. For all simulations, the hard core radius $R_{HD} =
9 \un{\AA}$ and the decay length $\xi = 15 \un{\AA}$.}
\label{fig:qmax}
\end{figure}

The effect of the additional potential on $S(q)$ is illustrated in
Figure~\ref{fig:qmax}, displaying the position $q_{\rm max}$ of the
first maximum in the structure factor as a function of pore density
for the Gram/DLPC system, as well as for simulated structure factors
with a hard core $R_{HD} = 9 \un{\AA}$ (the geometric core radius of
the pore) and an additional component (Equation~\ref{eq:vr}) with a
decay length $\xi = 15 \un{\AA}$ and various amplitudes $u$
(positive and negative).

The first observation is that, for a purely hard-core interaction,
the position of the maximum in $S(q)$ varies very little with the
concentration. Adding an attractive potential $u<0$ only shifts the
peak to higher $q$ values (the pores are on the average closer to
their neighbours) without changing its slope. When the potential is
repulsive $u>0$, on the other hand, the maximum shifts to lower $q$
values and its density dependence becomes steeper, although not
enough to describe the variation of the experimental data, even when
the parameters $u$ and $\xi$ are allowed to vary within reasonable
limits. A preliminary conclusion would thus be that the interaction
potential contains a ``soft'' repulsive contribution, induced by the
membrane, which varies with the density.

To check this conclusion, we made extensive attempts to fit the data
to a hard-core model plus an additional interaction (Equation
\ref{eq:vr}), where $R_{HD} = 9 \un{\AA}$ is the same for all
curves, and $n_{\rm pore}$ is fixed at the theoretically expected
value, given by the dotted line in Figure~\ref{fig:HD_fit}, while
$u$ and $\xi$ are allowed to vary freely. Alternative functional
forms for the additional interaction (exponential and linear) were
also tested. None of these attempts yielded satisfactory results; in
particular, parameter combinations that capture the density
dependence of the peak position $q_{\rm max}$ also give a marked
variation in its width and amplitude, in contrast with the
experimental data in Figure~\ref{fig:HD_fit}~a), where the shape of
the peak changes very little as $P/L$ varies by a factor of five.

\subsection{Possible origins for the interaction} \label{subsec:origin}

To see why and in what manner the membrane-mediated interaction
should vary with the pore density, we must first consider the effect
of inclusions on the membrane. Two such effects are relevant in this
context. The first one is related to changes in the membrane
thickness and occurs on a ``mesoscopic'' scale, affording a
continuum elasticity treatment. The second effect concerns the way
in which a membrane inclusion perturbs the configuration of the
lipid chains and requires a more involved, microscopic description.

\subsubsection{Hydrophobic matching} \label{subsubsec:mismatch}

One of the simplest (and surprisingly successful) concepts used to
interpret protein-membrane interaction is that of
\textit{hydrophobic matching} \citep{Mouritsen:1984,Killian:1998},
stating that proteins with a certain hydrophobic length (defined by
their transmembrane domain) are targeted to membranes with a
matching hydrophobic thickness. When there is a difference in length
between the hydrophobic part $h$ of the protein or peptide and that
of the host membrane (referred to as the ``hydrophobic mismatch''),
the bilayer deforms and adapts to the protein (which is generally
much more rigid). This deformation has a certain lateral extension
(in the plane of the membrane) and therefore induces an interaction
between inclusions when the latter are closer than this distance.

The hydrophobic length of gramicidin was estimated at $h_G = 22
\un{\AA}$ \cite{Elliott:1983}. For DLPC, $h = 20.8 \un{\AA}$ in the
pure membrane and $h = 22.1 \un{\AA} \approx h_G$ for a gramicidin
content $P/L = 0.1$ \cite{Harroun:1999}, confirming that thinner
membranes are stretched by gramicidin. The same effect is observed
in the gramicidin/DDAO system, where NMR measurements yield a
hydrophobic thickness of 18.4 {\AA} for the pure bilayer, increasing
to 19.4 {\AA} for the more concentrated samples ($P/L = 0.25$)
\cite{Oradd:1995}. For $\un{C_{12}E_{5}}$, we estimate the
hydrophobic thickness in the absence of gramicidin at
$h_{\un{C_{12}E_{5}}} = 18.8 \un{\AA}$; we have no data on the
variation of $h_{\un{C_{12}E_{5}}}$ with the pore density.

To first order (and ignoring the effect of the spontaneous curvature
of the monolayer, which can have non-trivial effects
\cite{Aranda:1996,Bohinc:2003}), the elastic energy is expected to
scale as $(h-h_G)^2$ \cite{Huang:1986}. It will therefore decrease
as the gramicidin concentration increases, and vanish when the
hydrophobic thickness of the membrane equals that of the bilayer.

\subsubsection{Changes in lipid ordering}
\label{subsubsec:order}

On the other hand, even in the absence of hydrophobic mismatch, the
presence of an inclusion restricts the configuration of lipid chains
in its vicinity
\citep{Marcelja:1976,Sintes:1997,Lague:2000,May:2000}. For instance,
Lag\"{u}e et al. \citep{Lague:2000,Lague:2001} used the lateral
density-density response function of the hydrocarbon chains obtained
from MD simulations of pure bilayers to determine the interaction
between ``smooth'' hard cylinders embedded in the bilayer. They
considered three values of the cylinder radius, up to 9~{\AA}, which
is precisely the geometric radius of the gramicidin pore. For the
largest radius, the long-range interaction is repulsive for all the
investigated lipids (DMPC, DPPC, POPC, DOPC), with an additional
short-range attraction in the case of DMPC. However, no specific
predictions are available for DLPC. Furthermore, it is not clear how
the interaction varies as a function of inclusion concentration.

\subsection{Complete model} \label{subsec:compmod}

In the following, we adopt the hydrophobic matching model and assume
that the mismatch $(h-h_G)$ varies linearly with the in-plane
concentration of pores $n_{\rm pore}$ over the investigated range
($1/25 \leq P/L \leq 1/5$, or $n_{\rm min} = 0.55 \leq n_{\rm pore}
\leq n_{\rm max} = 1.7 \times 10^{-3}~\un{\AA}^{-2}$) and vanishes
at the highest value. We model this elastic interaction by an
exponential\footnote{We also performed tests using different
functional forms for the interaction potential, but the fits are
less satisfactory (see the appendix).}
\begin{equation}
V(r)=U_0 \left ( \frac{n_{\rm max} - n_{\rm pore}}{n_{\rm max} -
n_{\rm min}} \right )^2 \exp \left ( -\frac{r - 2 R}{\xi} \right )
\label{eq:interact}
\end{equation}
where the free parameters are the prefactor $U_0$, which is also the
interaction amplitude at the lowest concentration $n_{\rm min}$, and
the decay length $\xi$, taken as independent of the concentration.
No effort was made to include three-body interactions. Since the
additional potential vanishes at the highest pore concentration, in
the following we take as hard core radius the best fit at this
value, namely 9.5 \AA, rather than the geometric value of 9.0 \AA.
This adjustment is not very significant, but is required for
coherence of the model and it ensures that the additional component
is not overestimated.

For the Gram/DLPC system, the best fit with this model is obtained
for $U_0= 31.5 \pm 10 \, k_B T$ and $\xi = 2.5 \pm 0.5 \un{\AA}$.
The details are given in Figure \ref{fig:final_fit}. See the
Appendix A for details on the error estimate.

\begin{figure*}
\includegraphics[width=0.7\textwidth]{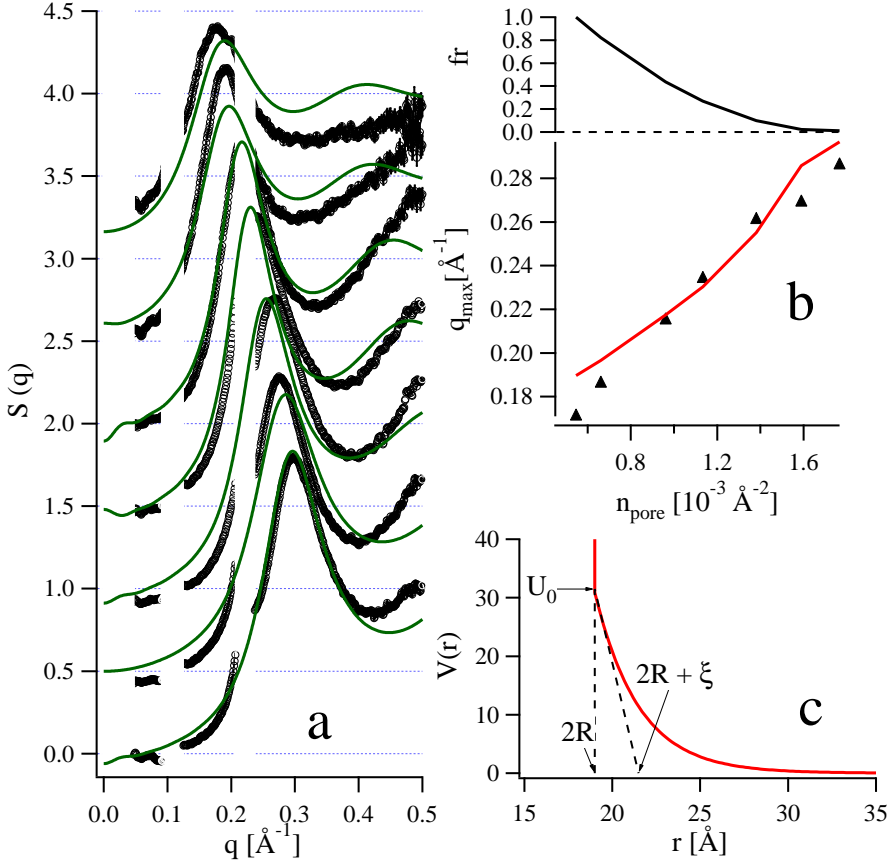}
\caption{(a) Experimental structure factors (symbols) and fits
(solid lines) with a hard-disk plus the repulsive interaction
(\ref{eq:interact}) for the gramicidin/DLPC system (data as in
Figure \ref{fig:HD_fit}). (b) Top: ``Effective fraction'' of the
interaction amplitude as a function of $n_{\rm pore}$: $fr=\left (
\frac{n_{\rm max} - n_{\rm pore}}{n_{\rm max} - n_{\rm min}} \right
)^2$, see Equation (\ref{eq:interact}). Bottom: Position of the
structure factor maximum $q_{\rm max}$ vs. $n_{\rm pore}$, for the
experimental data (symbols) and for the fits (solid line). (c)
Interaction potential $V(r)$ used for the fits in panel (b). The
amplitude corresponds to the lowest value of $n_{\rm pore}$, $n_{\rm
min} = 0.55 \times 10^{-3}~\un{\AA}^{-2}$.} \label{fig:final_fit}
\end{figure*}

\begin{figure}
\includegraphics[width=0.5\textwidth]{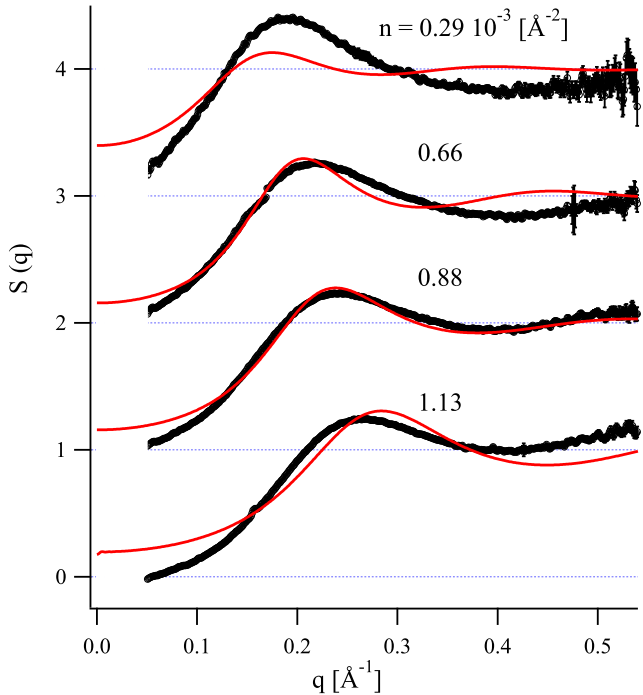}
\caption{Experimental structure factors (symbols) and fits (solid
lines) with a hard-disk plus the repulsive interaction
(\ref{eq:interact}) for the gramicidin/C$_{12}$EO$_{5}$ system. The
pore density is indicated alongside each curve (the four curves
correspond to the solid squares in Figure \ref{fig:HDradius}).}
\label{fig:C12E5}
\end{figure}

A similar result is obtained for the Gram/C$_{12}$EO$_{5}$ system:
$U_0= 27 \pm 10 \, k_B T$ and $\xi = 2.75 \pm 0.5 \un{\AA}$ (see
Figure \ref{fig:C12E5}). As above, the mismatch $(h-h_G)$ is taken
to vary linearly from a maximum at the lowest pore concentration to
zero in the most concentrated sample.

For the Gram/DDAO system, fitting attempts point towards a much
lower membrane-mediated interaction (both in amplitude and decay
length), coherent with the low variation in effective hard disk
radius (open dots in Figure \ref{fig:HDradius}.) Clearly, the
interaction of the pores is closer to a pure hard-disk model than
for the other bilayer compositions. However, a quantitative analysis
is difficult in view of the low quality of the fits. Further
experimental data is needed on this system.

\section{Conclusion} \label{sec:conc}

We showed that the interaction between gramicidin pores inserted
within DLPC bilayers can be well described by a hard disk potential,
with a range close to the geometric diameter of the molecule and an
additional repulsive interaction, whose amplitude decreases with
increasing pore concentration. A similar result is obtained for
pores inserted within C$_{12}$EO$_{5}$ bilayers.

The decrease of the interaction amplitude with pore concentration is
consistent with the hydrophobic matching hypothesis, whereby the
increase in bilayer thickness (well-documented for Gram/DLPC systems
\cite{Harroun:1999}) helps ``accommodate'' the peptide and leads to
a decreased interaction. The decay length of the interaction is
found to be of the order of 2.5~\AA, well below the value predicted
by continuum elastic models \cite{Aranda:1996,Harroun:1999b} and
close to the characteristic length given by more microscopic models
\cite{Fattal:1993}. However, considering that the amplitude of the
interaction is quite high, the interaction range --defined
intuitively as the distance over which the pores `see' each other--
does extend over several decay lengths; see Appendix \ref{sec:comp}
for a more detailed comparison.

An intriguing consequence of the interaction decreasing with the
concentration of pores is that the structure factors of the
two-dimensional fluid that they form in the plane of the membrane
changes relatively little with the area density (as compared to e.g.
purely hard-core interactions). One could say that the degree of
``liquid order'' remains sensibly the same. It is tempting to
speculate on the biological significance of this feature. For
instance, does the activity of gramicidin require a certain degree
of correlation between pores? We hope that systematic studies of the
interaction in various conditions will help answer this question.

\appendix

\section{Data fitting}

The first observation is that the (statistical) uncertainty in the
structure factor $S(q)$, determined by radial regrouping, background
subtracting and then dividing by the calculated form factor, is much
too small. Clearly, the discrepancy between the fit functions and
the data is mostly due to systematic effects, which can be related
to errors in determining $S(q)$ from the measured intensity or to
the inadequacy of the fit functions. We therefore used a more
realistic uncertainty value, $\sigma=0.1$ taken as a constant for
all data points. The goodness-of-fit function $\chi ^2$ is then of
the order of 2 (per data point.)

\begin{figure}
\includegraphics[width=0.5\textwidth]{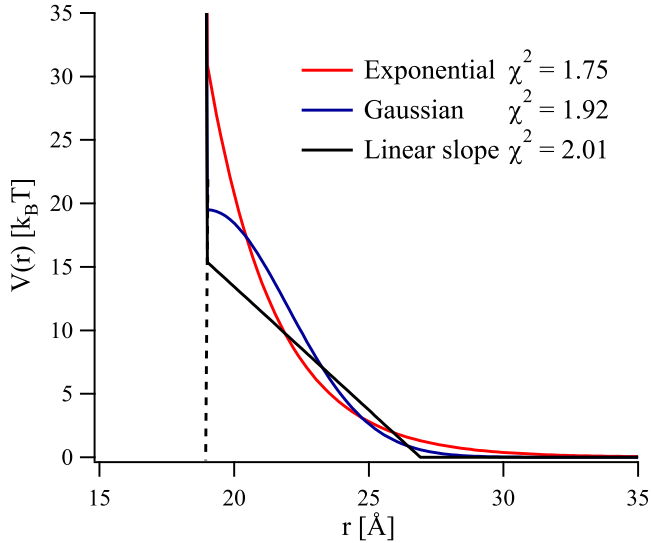}
\caption{The best results for the interaction potential $V(r)$
within each model class. The respective $\chi ^2$ values are also
indicated.\label{fig:comp}}
\end{figure}

Several trial functions for the interaction potential $V(r)$ were
tested: exponential decrease, gaussian function and linear slope.
The exponential model presented in Figure \ref{fig:final_fit} yields
the best fit. The best potentials obtained for each class of model
are displayed in Figure \ref{fig:comp}. The respective $\chi ^2$
values are indicated alongside the curves.

\begin{figure}
\includegraphics[width=0.5\textwidth]{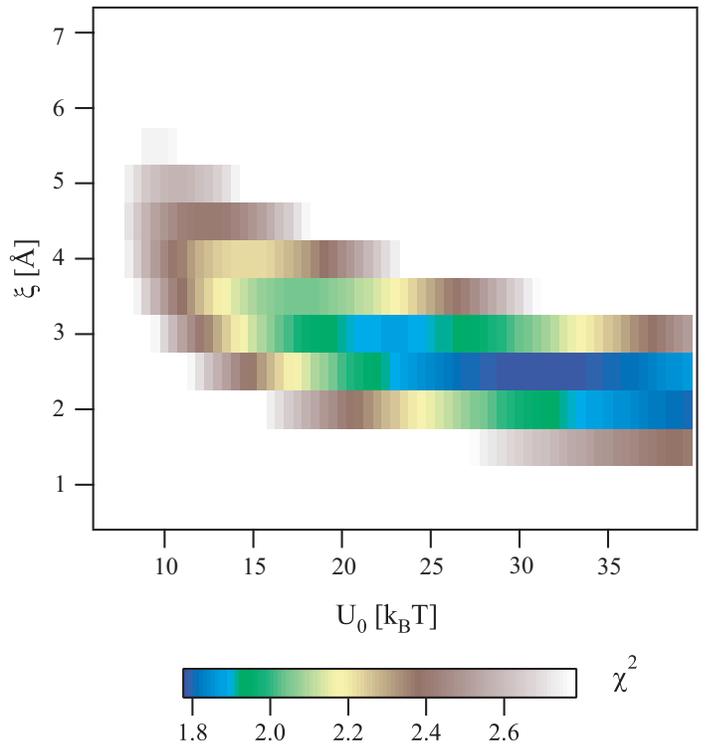}
\caption{Goodness-of-fit function $\chi ^2$ as a function of the
parameters $U_0$ and $\xi$.\label{fig:chisq}}
\end{figure}

Once the fit is obtained, the confidence intervals should be
determined for each parameter. For the exponential model, which
yields the best fit, we computed $\chi ^2 (U_0, \xi)$ for a wide
range of parameters, to make sure that we have indeed found the
global minimum and to visualise the dependence of $\chi ^2$ on the
parameters. This graph is plotted in Figure \ref{fig:chisq}.
Clearly, $\xi$ is within the range $2.5 \pm 0.5 \un{\AA}$. On the
other hand, the interaction amplitude is less well defined; we take
the conservative estimate $U_0= 31.5 \pm 10 \, k_B T$.

\section{Comparison of the interaction range}
\label{sec:comp}

\begin{figure}
\includegraphics[width=0.5\textwidth]{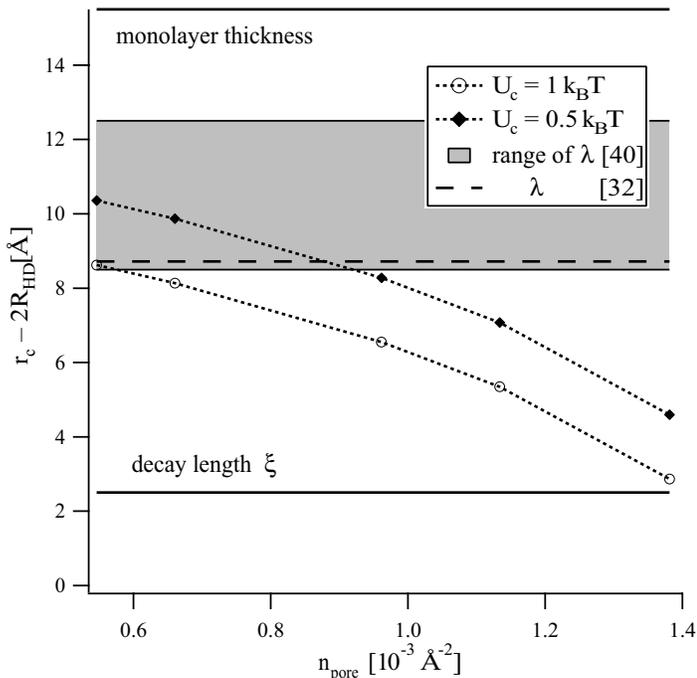}
\caption{Distance scales relevant for the interaction: the decay
length $\xi$ measured in this paper and the monolayer thickness
(both shown as solid lines), the cutoff distances $r_c$ for two
cutoff amplitudes, 0.5 and $1~k_BT$ (symbols and dotted lines) and
the values of the characteristic length $\lambda$ as given by
\cite{Harroun:1999b} (gray shaded range) and estimated from
\cite{Aranda:1996} (dashed line). \label{fig:range}}
\end{figure}

As mentioned in the Conclusion, $\xi$ is not entirely
representative, since the interaction amplitude can remain
significant over several decay lengths. This is an obstacle to
comparing the different theoretical models between them and with the
experimental data, unless they can be described by a common
functional form. Another --less precise-- option is to define an
arbitrary cutoff amplitude and to define the interaction range as
the separation beyond which the interaction falls below this cutoff.
We did this for our model with two cutoff values, $U_c = 0.5$ and
$1~k_BT$, respectively. The resulting cutoff ranges $r_c$ are shown
in Figure \ref{fig:range} (symbols and dotted lines). For reference,
the decay length $\xi$ is also plotted as solid line.

In continuum elasticity models, the characteristic length is
$\lambda = \left ( \frac{hK}{4B} \right )^{1/4}$
\cite{Harroun:1999b}, with $K$ and $B$ the bending and compression
moduli of the bilayer, respectively. These authors conclude that a
characteristic length in the range $8.5 \leq \lambda \leq 12.5
\un{\AA}$ (shown in Figure \ref{fig:range} as gray shading) accounts
for the behaviour of gramicidin in DMPC bilayers. We also plotted as
dashed line the characteristic length used by Aranda-Espinoza et al.
\cite{Aranda:1996} (corresponding to $\beta = 10$ in their choice of
parameters). Clearly, these values are much larger than our value
for the decay length, $\xi = 2.5 \un{\AA}$, with $\lambda / \xi$
between 3 and 5.

In order to compare the effective interaction range, one would need
to estimate the cutoff ranges for the theoretical results. Although
this data is not available in the publications, it appears that the
interaction extends at least over the thickness of one monolayer
(shown in Figure \ref{fig:range} as solid line), which is also
quoted as the dominant scale length in reference \cite{Dan:1993}.
This length scale is clearly larger than our estimates for the
cutoff range, although the discrepancy is less marked than between
$\lambda$ and $\xi$.

\section*{Acknowledgements}

The ESRF is gratefully acknowledged for the provision of synchrotron
radiation facilities (experiment 02-01-732) and we thank C. Rochas
for competent and enthusiastic support. B. deGroot is acknowledged
for providing the atomic coordinates of the gramicidin channel and
an anonymous referee for suggesting the analysis in Appendix
\ref{sec:comp}.


\bibliography{ala_ref,gram}

\end{document}